\documentclass[%
 reprint,
 amsmath,amssymb,
 aip,
floatfix,
]{revtex4-1}

\usepackage{graphicx}
\usepackage{dcolumn}
\usepackage{bm}
\usepackage{color}
\usepackage{hyperref}
\usepackage{xcolor,soul}
\hypersetup{
    colorlinks=true,
    linkcolor=blue,
    citecolor=blue,
    urlcolor=cyan,
}
\usepackage{natbib}
\usepackage[utf8]{inputenc}

\usepackage[normalem]{ulem}

\begin{document}

\title{Voltage control of frequency, effective damping and threshold current in nano-constriction-based spin Hall nano-oscillators}

\author{Victor H. Gonz\'alez}
\author{Roman Khymyn}%
\affiliation{Physics Department, University of Gothenburg, 412 96 Gothenburg, Sweden.}
\author{Himanshu Fulara}
\affiliation{Department of Physics, Indian Institute of Technology Roorkee, Roorkee 247667, India}
\author{Ahmad A. Awad}
\author{Johan {\AA}kerman}
\email{johan.akerman@physics.gu.se}
\affiliation{Physics Department, University of Gothenburg, 412 96 Gothenburg, Sweden.}

\date{\today}

\begin{abstract}
Using micromagnetic simulations, we study the interplay between strongly voltage-controlled magnetic anisotropy (VCMA), $\Delta K = \pm$200 kJ/m$^3$, and gate width, $w=$ 10--400 nm, in voltage-gated W/CoFeB/MgO based nano-constriction spin Hall nano-oscillators. The VCMA modifies the local magnetic properties such that the  magnetodynamics transitions between regimes of \emph{i}) confinement, \emph{ii}) tuning, and \emph{iii}) separation, with qualitatively different behavior. We find that the strongest tuning is achieved for gate widths of the same size as the the constriction width, for which the effective damping can be increased an order of magnitude compared to its intrinsic value. As a consequence, voltage control remains efficient over a very large frequency range, and subsequent manufacturing advances could allow SHNOs to be easily integrated into next-generation electronics for further fundamental studies and industrial applications. 

\end{abstract}

\maketitle

Spin Hall nano-oscillators (SHNOs)~\cite{liu2012prl,demidov2012ntm, liu2013prl,demidov2014ntc,demidov2014apl,duan2014ntc,Sato2019PRL} are miniaturized ultra-broadband microwave signal generators, typically based on bilayer stacks of nonmagnetic heavy-metal (HM) and ferromagnetic (FM) thin films, where the spin Hall effect~\cite{hirsch1999prl, hoffmann2013ieeem, sinova2015rmp} 
converts a direct longitudinal current in the HM into a pure transverse spin current injected into the FM, where the associated spin transfer torque~\cite{slonczewski1996jmmm,berger1996prb,ralph2008jmmm}  can drive auto-oscillations of the local magnetization. Nanoconstriction-based SHNOs~\cite{demidov2014apl,Awad2016NatPhys,Durrenfeld2017Nanosc,Mazraati2016apl,Zahedinejad2018APL,mazraati2018pra,Awad2020APL} have a variety of desirable properties, such as robust RF response~\cite{Awad2020APL}, ease of injection locking~\cite{Hache2019APL} and mutual synchronization~\cite{Awad2016NatPhys,Zahedinejad2020natnano}, making them ideal candidates for broadband signal processing and unconventional oscillator computing, where multi-signal injection locking has shown potential for neuromorphic computing~\cite{yogendra2016ieeeted,romera2018nature,Zahedinejad2020natnano,mcgoldrick2022ising}. SHNOs are also potentially appealing for oscillator network-based Ising machines to solve combinatorial optimization problems.\cite{albertsson2021ultrafast, houshang2022phase}

To unlock their potential for computing, effective control of the functional properties of each individual SHNO must be implemented. While optothermal SHNO control was very recently experimentally demonstrated~\cite{MuralidharAPL2022}, a more established approach has been to use gates and voltage controlled magnetic anisotropy (VCMA) to tune the material properties of the active region of the SHNO, and, through this, its microwave signal properties. Back-gating a nanogap SHNOs achieved about 4.6~MHz/V of frequency tuning~\cite{liu2017PRA}. In nanoconstriction SHNOs, a frequency tuning of similar order of magnitude was recently demonstrated for  CoFeB sandwiched between W and MgO~\cite{Fulara2020}, which, thanks to the interplay between the device geometry and the VCMA, also resulted in a giant change in the effective damping and the SHNO threshold current. By adding a memristive behavior to the gate dielectric, the tuning could also be made non-volatile.\cite{zahedinejad2022natmat}

So far, only a relatively limited tuning range has been explored. Very recently, Choi et al, presented giant SHNO frequency tuning of up to 2.1 GHz using 
VCMA tuning in Co/Ni multilayers, with a reported a change of the perpendicular magnetic anisotropy (PMA) as large as 28.2~kJ/m$^3$V.~\cite{Choi2022natcomm} This hence raises the question of how the damping, threshold current, and overall auto-oscillation behavior would behave if such strong tuning could be achieved with narrow gates, where geometrical effects should further amplify the effect. 

\begin{figure}[b]
    \centering
    \includegraphics[width=7.7cm]{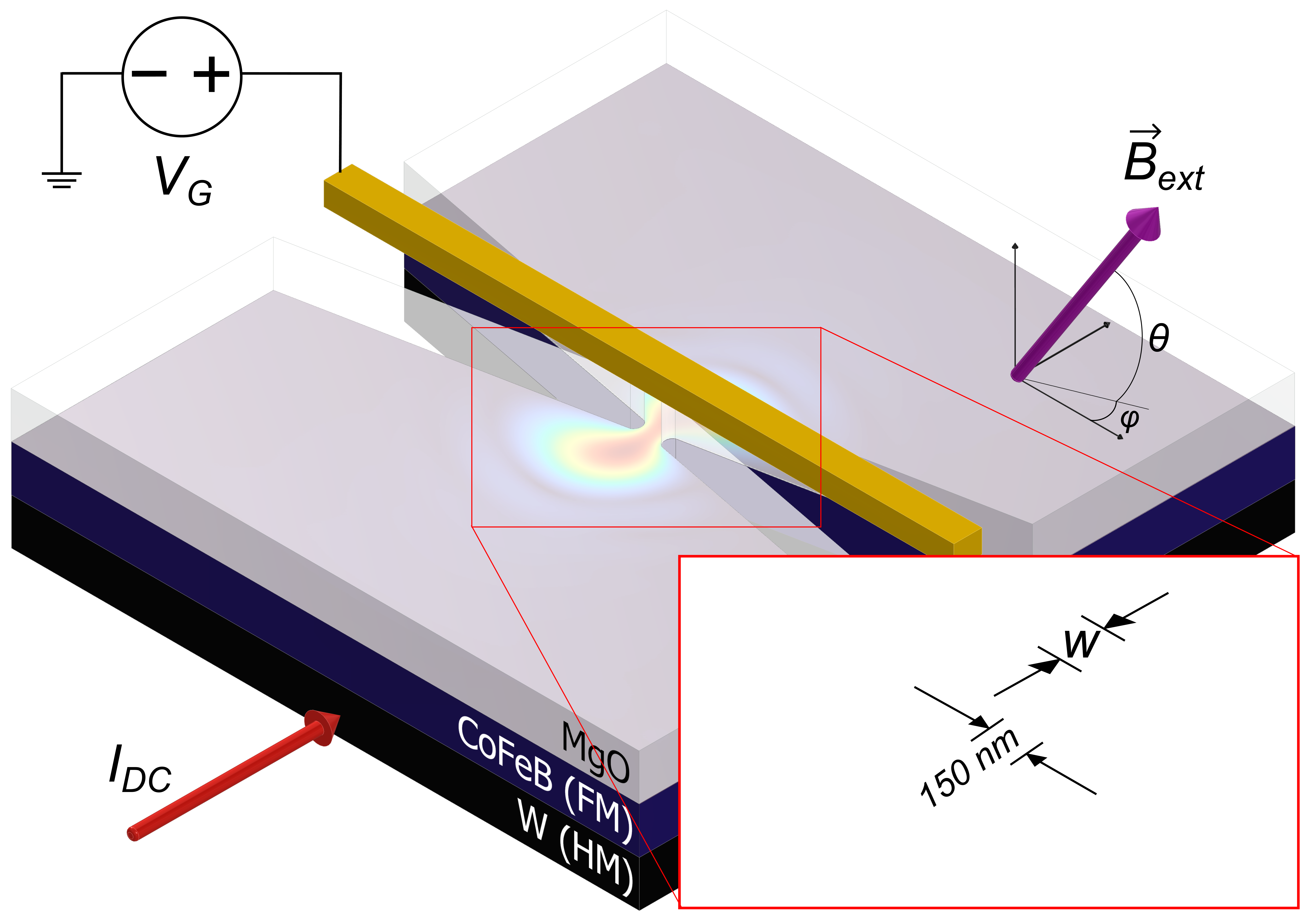}
    \caption{Schematic of the simulated voltage-controlled  W(5~nm)/CoFeB(1.7~nm)/MgO(2~nm) SHNO.}
    \label{fig:layout_gated-shno}
\end{figure}

Here, we use micromagnetic simulations to study voltage-gated SHNOs over a very wide range of PMA tuning ($\Delta K = \pm$200 kJ/m$^3$) and probe its effect on the damping, the resonant frequency, and the excited modes for a wide range of gate widths ($w=$ 10--400 nm). We find a broad range of tunability for both frequency and damping due to changes in mode localization both depending on voltage and gate width. Our results demonstrate the versatility of VCMA in SHNOs and suggest new phenomena, not yet seen in experiments, which could be employed in the efficient design of purpose-built devices for low-power electronics and unconventional computing applications.

We simulated a W(5~nm)/CoFeB(1.7~nm)/MgO(2~nm) bow tie shaped bilayer with a nanoconstriction (NC) of 150~nm width, curvature radius of 50~nm, and opening angle of 22$^\circ$, as shown in Fig.~\ref{fig:layout_gated-shno}. The geometry of the sample is taken from experiments in earlier work.\cite{Fulara2020} The in-plane current density and the Oersted field were sampled using the COMSOL multiphysics software~\cite{COMSOL}. The spin-polarized current density acting on the CoFeB layer due to the spin Hall effect was calculated by multiplying the in-plane current density at the W layer by a spin-Hall angle $\theta_\mathit{{SH}}=0.43$ and the injected spin current polarized perpendicular to the local calculated current density. 

The geometry of the SHNO, with a total volume of 4096$\times$4096$\times$1.7~nm$^3$, was discretized into 1024$\times$1024$\times$1 cells and loaded into the GPU-accelerated micromagnetic simulator Mumax3~\cite{vansteenkiste2014aip}. The properties of the CoFeB layer were chosen as: 
exchange stiffness $A_{ex}=19$ pJ/m, saturation magnetization $M_{S}=1050$ kA/m, Gilbert damping constant $\alpha_0=9\times 10^{-3}$, and gyromagnetic ratio $\gamma/2\pi= 29.1$ GHz/T. An oblique magnetic field of $B_{ext}=$ 0.4T was applied at polar $\theta=$ 60.5$^{\circ}$ and azimuthal $\phi=$ 22.0$^{\circ}$ angles consistent with previous experiments.\cite{Fulara2020} To reduce artifacts due to spin wave reflections from the edge, periodic boundary conditions were used with absorbing borders, in the form of quadratically interpolated increased damping.

The VCMA was applied to a rectangular gated region by a voltage-dependent uniaxial perpendicular magnetic anisotropy (PMA) of the form,
\begin{equation}
    K=645 \frac{kJ}{m^3} - V_G\cdot 2.5 \frac{kJ}{m^3 V},
    \label{eq:VCMA}
\end{equation}
where both the intrinsic $K(V_G=0)$ and its linear gate voltage dependence 
were taken from experiments.\cite{Fulara2020} $V_G$ was swept from --80~V to +80~V to investigate a much wider range than studied previously. 
Although such high voltages are unlikely to be used, the corresponding change in $K$ can be realized at lower voltages using materials with higher VCMA coefficients.\cite{Choi2022natcomm}

\begin{figure}[b]
    \centering
    \includegraphics[width=7.8cm]{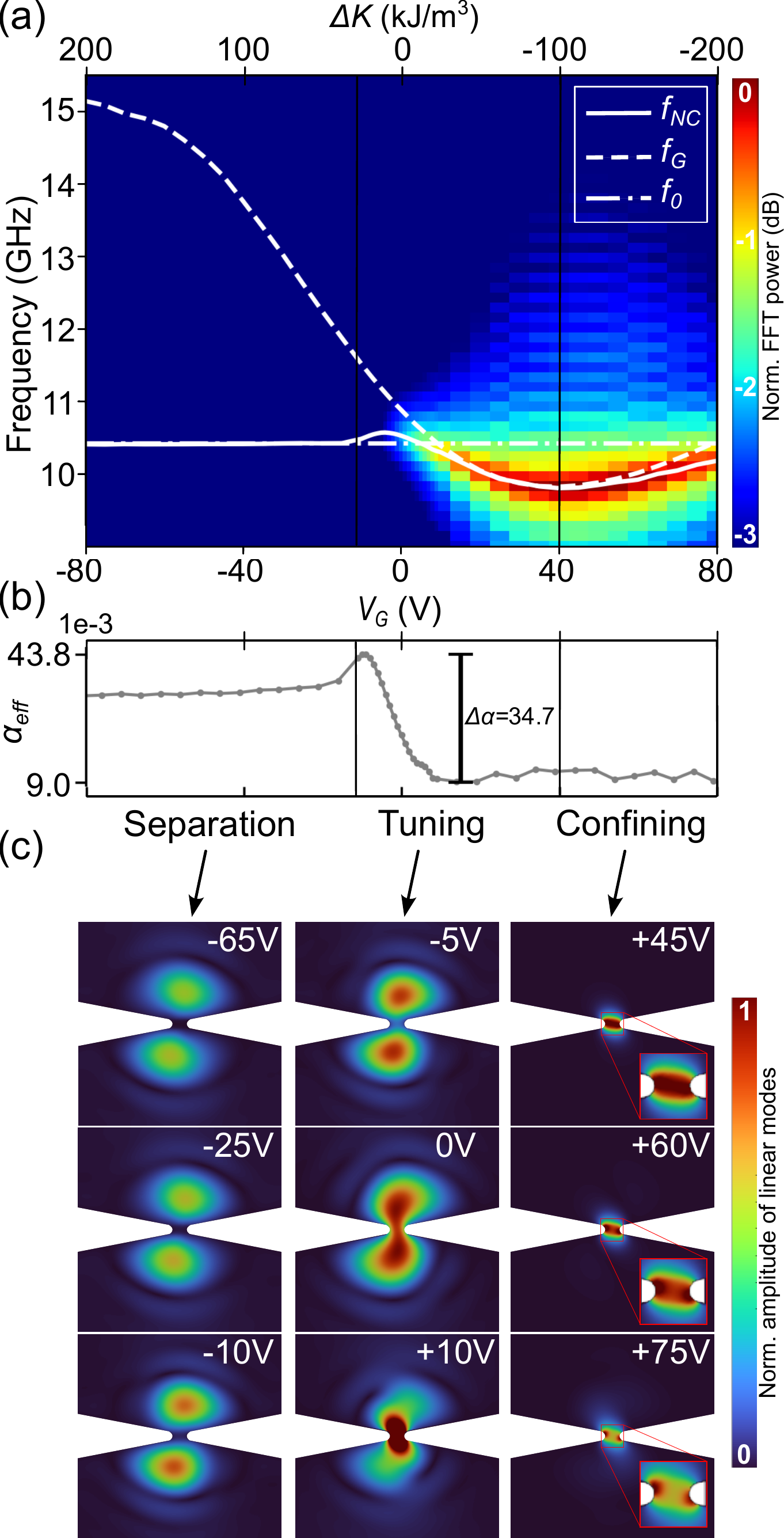}
    \caption{Linear response of a voltage-gated SHNO with 96~nm of gate width. (a) PSD (color map) and extracted frequencies (white lines)  and (b) Effective damping ($\alpha_\mathit{{eff}}$), as functions of gate voltage for the pulsed current ($V_\mathit{G}$) and the change in PMA ($\Delta K$). (c) Amplitude profiles of the magnetodynamical modes for selected gate voltages showing the features of the three gating regimes, confinement, tuning, and separation (black lines in (a,b). The inset zoom-in around the constriction region.}
    \label{fig:96nm}
\end{figure}

For extraction of the linear response of the system, such as the dominant resonances and their damping, 
the FM layer was first relaxed, then excited with a short Gaussian pulse ($\sigma =$ 10~ps) and allowed to decay back to equilibrium for 5~ns with a sampling rate of 10~ps. 
Then, the average magnetization in the center of the gate was converted to the frequency domain via a fast Fourier transform (FFT) and then fitted with a Lorentz distribution centered on the FMR frequency $f$ and with linewidth $\Delta f$. 
From this fitting, the effective damping constant was calculated as
\begin{equation}
    \alpha_\mathit{{eff}}=\frac{\gamma\mu_0}{2\pi}\frac{\Delta f}{f}\frac{dH}{df}
\end{equation}
for all $V_G$.\cite{Fulara2020} The average power was also estimated using the relation $P\propto \mathcal{F}(m)^2$, where $\mathcal{F}(m)$ is the Fourier transform of the magnetization, to study all magnetodynamic modes. 

The power spectral density (PSD) map shown in Fig.~\ref{fig:96nm}(a) was constructed as a function of applied voltage and a gate of 96~nm. The nanoconstriction frequency, $f_\mathit{{NC}}$,  and $\alpha_\mathit{{eff}}$ (Fig.~\ref{fig:96nm}(b)) were extracted for each case. Both quantities show a non-monotonic behavior and wide ranges of tunability, with a five-fold increase in damping within the -10~V to +10~V range and an accessible band of frequencies of approximately 800~MHz  within the -5~V to +40 V range. 

To investigate the origin of the nonmonotonic frequency behavior of the constriction mode, we simulated a FM layer with a geometry restricted to the gated region. The extracted frequency of this gate is plotted in Fig.~\ref{fig:96nm}(a), along with the intrinsic frequency $f_0$ of an ungated NC. From comparing the three frequency curves, we can conclude that the shape of $f_\mathit{{NC}}$ emerges from the interplay of an intrinsic mode $f_0$ and a gated non-linear mode $f_G$. In principle, we can switch between these two modes with voltage alone.

To gain more insight into the VCMA-mediated modulation of the effective damping and dynamics of the SHNO, we compared the spatially resolved mode profile amplitudes at $f_\mathit{{NC}}$ for all $V_G$. The mode profiles were constructed through a point-wise FFT, performed on magnetization snapshots sampled at 10 ps intervals, producing both the amplitude and phase maps of the oscillation modes. The mode amplitudes are shown in Fig.~\ref{fig:96nm}(c) where three voltage gating regimes can be identified based on the effect VCMA has on their features: a separation regime, a tuning regime, and a confining regime.

For strong negative voltages, the effective PMA under the gate becomes large and suppresses the excitation of localized modes in the gate region. Thus, the modes delocalize and separate into two coupled in-phase regions, as shown in the left column of Fig.~\ref{fig:96nm}(c). This \emph{separation regime} spans from approximately --80~V to --10~V (or $\Delta K$ 200~kJ/m$^3$ to 12.5~kJ/m$^3$) and presents mostly flat $\alpha_\mathit{{eff}}$ and $f_\mathit{{NC}}$ curves. Because the oscillations are driven out, the NC is driven mostly by the evanescent modes outside of it, which implies that the excited spin-wave volume increases, promoting radiative damping~\cite{slavin2005prl, demidov2012ntm}. The volume of this mode remains approximately the same and thus the damping is constant with voltage.

The \emph{tuning regime}, with gate voltage around roughly --10~V to +40~V ($\Delta K$ 12.5~kJ/m$^3$ to --100~kJ/m$^3$), is where we observe the largest tunabilities. From the middle column of Fig.~\ref{fig:96nm}(c), this regime shows the biggest change in excited mode volumes, which explains 
the 800~MHz tunability of frequency and five-fold increase in effective damping. These are due to the volume delocalization of the excited mode, which pushes emitted magnons into the non-gated region while preserving the mode's core. The flattening of $\alpha_\mathit{{eff}}$ above +15V comes from the fact that the oscillations are limited to the gated region and we recover the intrinsic $\alpha_0$. Also, in this regime, the non-linear gated mode is close with the intrinsic mode of the system and thus drives the overall dynamics, reshaping $f_\mathit{{NC}}$. Such a range lends the device a lot of versatility as a microwave component, with the advantage of the low-power consumption a voltage source permits.

For high positive voltages between +40~V to +80~V ($\Delta K$ --100~kJ/m$^3$ to --200~kJ/m$^3$),
the local out-of-plane effective field angle is smoothly lowered with larger voltage (and smaller PMA). In this interval, the excited modes localized within the constriction, this regime also shows a minimum and a change in the sign of the derivative of $f_\mathit{{NC}}$. In this \emph{confining regime}, the localized mode shifts between the bulk (where the oscillations occur mainly in the middle of the constriction) and the edge (where they occur mainly at the edges)~\cite{Dvornik2018}, as shown in the right column of Fig.~\ref{fig:96nm}(c). 

\begin{figure}[t]
    \centering
    \includegraphics[width=8cm]{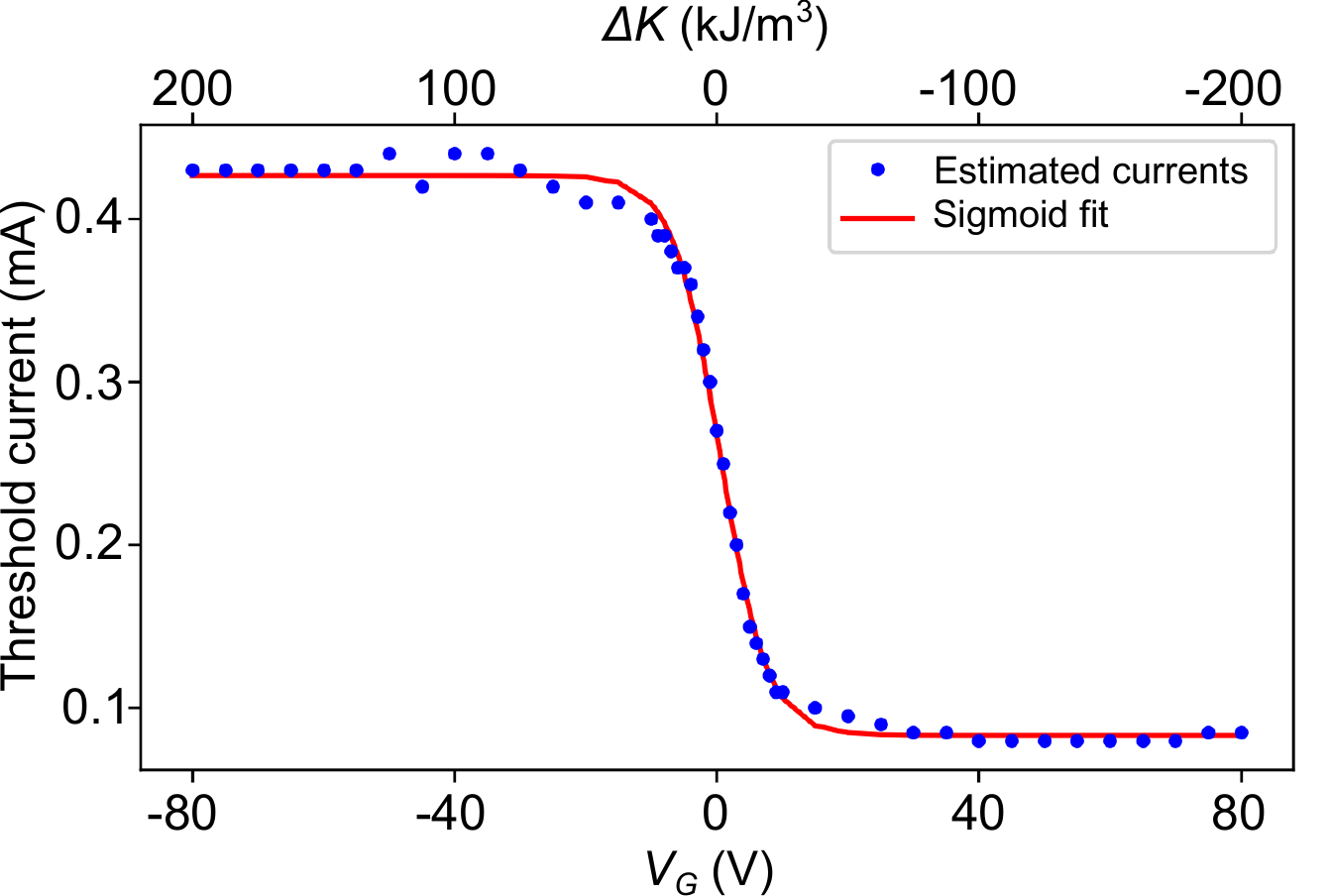}
    \caption{Threshold current as a function of gate voltage (change of PMA top axis) fitted with a sigmoid logistic function as a guide to the eye.
    }
    \label{fig:Ith-vs-VG}
\end{figure}

To determine the auto-oscillation threshold current, 
direct currents between 10 and 400~$\mu$A were applied and analyzed
for all $V_G$. The auto-oscillation threshold is defined as the smallest current producing an exponentially growing amplitude. Fig.~\ref{fig:Ith-vs-VG} shows the auto-oscillation threshold currents as a function of $V_G$. The shape of the curve  
resembles a sigmoid logistic function, as shown in the fit, and 
it follows the behavior of the damping shown in Fig.~\ref{fig:96nm}(b), as expected. The sigmoid behavior draws similarities with an activation function in an artificial neural network, which invites exploration of possible technological avenues for arrays of SHNOs based on voltage-based synaptic activation. Very recent work seems to suggest a voltage-enabled synaptic potentiation and depression that could permit SHNO based neurons without external memory circuitry~\cite{Choi2022natcomm}.

\begin{figure}
    \centering
    \includegraphics[width=1\linewidth]{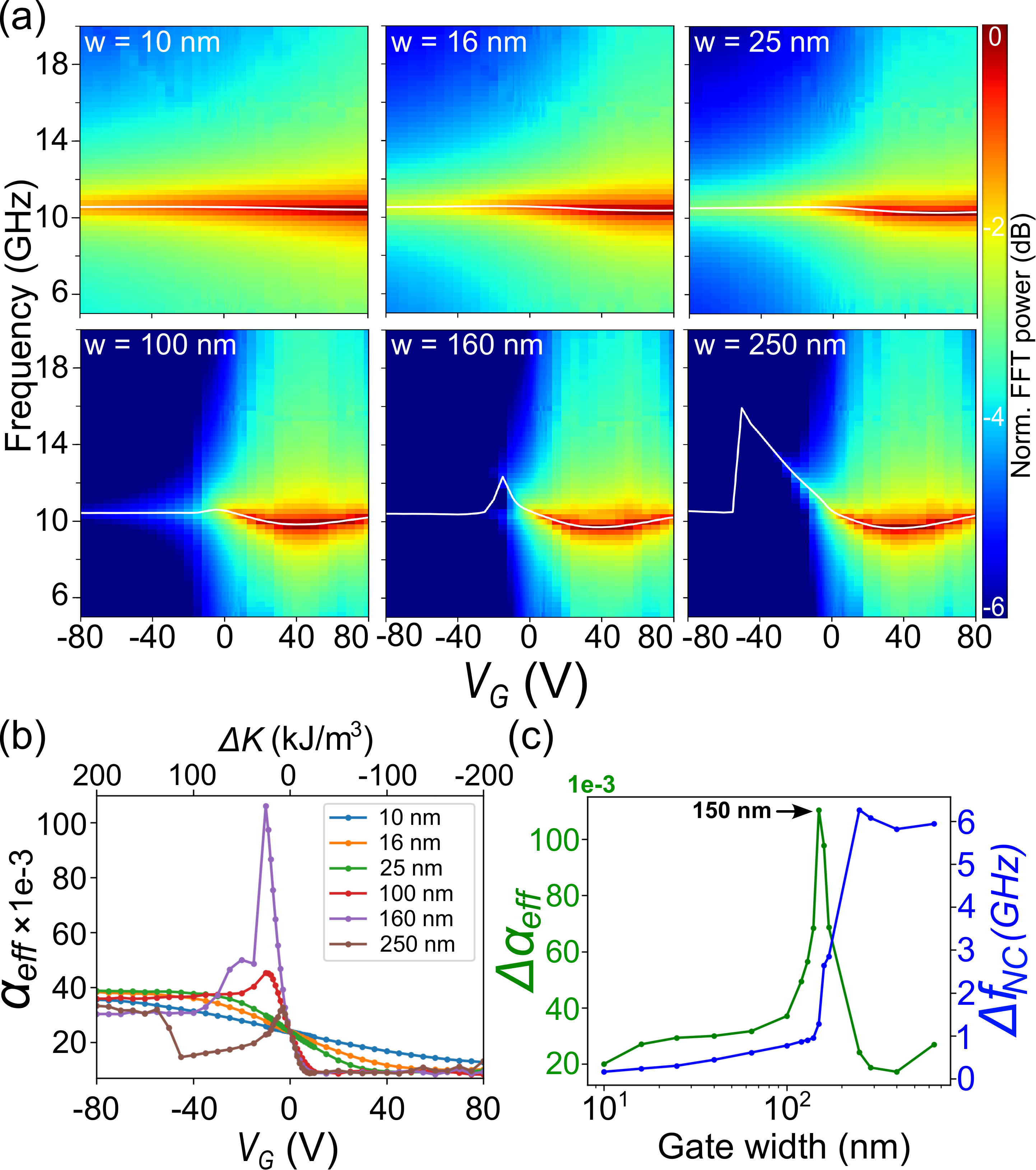}
    \caption{(a) PSD and extracted frequencies for six different gate widths, with the bottom gates being ten times wider than the top ones.
    (b) Effective damping 
    for the gates depicted in (a). 
    (c) Damping and frequency tunabilities as a function of gate width. Radiative damping shows a maximum $\Delta\alpha_\mathit{{eff}}$ at 150 nm.
    }
    \label{fig:different-gate-widths}
\end{figure}

Finally, to investigate the effect of the gate size on the modes, simulations and calculations of the frequency, damping, and power were repeated for a wide range of gate widths between 10~nm and 400~nm.
In Fig.~\ref{fig:different-gate-widths}(a), where we show PSDs for six widths, 
a wider gate improves the tunability of the frequency by affecting (either by localization or delocalization) a larger area, thus increasing the concavity of the gated mode.
For the three widest gates at the bottom, we can lower the voltages at which the gated mode is observable before the system switches back to its intrinsic mode. Furthermore, at w = 160~nm, we can see mode coexistence in the 0~V to --25~V interval.

Fig.~\ref{fig:different-gate-widths}(b), shows that the width of the gate plays an important role in the radiative damping process. As the width of the gate increases, there is a gradual change in the shape of $\alpha_\mathit{{eff}}$, with a sudden order of magnitude increase at 160 nm and a subsequent reduction at higher values. For this width value, $\alpha_\mathit{{eff}}$ increases with PMA up to a point, where PMA compensates the local demagnetization (at $-10$ V), i.e. the sharp maximum is observed at the transition from in-plane to out-of-plane type of anisotropy.

The damping ($\Delta \alpha_\mathit{{eff}}$) and frequency ($\Delta f_{NC}$) tunabilities are plotted in Fig.~\ref{fig:different-gate-widths}(c) as functions of the gate width. $\Delta f_{NC}$ grows monotonically as the PMA is compensated in a bigger gated area. However, around 150~nm, there is a large increase in the slope of the curve before it saturates for wider gates. 150~nm corresponds to the constriction size and it is the width at which full compensation of the demagnetization field is achieved; wider gates lead to saturation of the compensation as the extending the area further does not affect compensation. There is also a sharp maximum in $\Delta\alpha_\mathit{{eff}}$ with a maximum at 150~nm gate width, 
where a matching between the width of the gate and the size of the mode's core  at the center of the nanoconstriction enhances the power transfer between the gated and non-gated regions and allows for a higher voltage-driven manipulation of the damping. For wider gates, the modes are driven out completely, and we have two uncoupled regions, with minimal damping tunability. The curves in Fig.\ref{fig:different-gate-widths}(c) could serve as a guide to design tailored devices for applications where a trade off between damping and frequency tunabilities might be important.

In conclusion, we find that the VCMA separates the local magnetodynamic properties of nanoconstriction-based SHNOs into three major regimes of mode confinement, tuning, and separation, with qualitatively different behavior. We find that the strongest tuning happens when the gate and nano-constriction have similar sizes, for which the damping can be varied by one order of magnitude and the frequency by several GHz. VCMA-oriented design paired with further developments in  heterostructure fabrication techniques can hence pave the way towards more complex SHNO architectures for next generation unconventional spintronic  computation platforms.

\section*{Acknowledgment}
This work was supported by the Swedish Research Council (VR) and the Horizon 2020 research and innovation programme (ERC Advanced Grant No.~835068 "TOPSPIN").

\section*{Data Availability Statement}
The code and output data that support the findings of this study are available from the corresponding author upon reasonable request.
\bibliography{apssamp}

\end{document}